\documentclass[%
 reprint,
longbibliography,
superscriptaddress,
 amsmath,amssymb,
 aps,
prl,
]{revtex4-1}

\usepackage{hyperref}

\usepackage{graphicx}
\usepackage{dcolumn}
\usepackage{bm}

\usepackage{ulem}
\usepackage{color}

\begin{document}
\title{Shortcuts to Adiabaticity with Inherent Robustness and without Auxiliary
Control}
\author{Yiyao Liu}
\affiliation{Guangdong Provincial Key Laboratory of Quantum Engineering and Quantum
Materials, School of Physics and Telecommunication Engineering, South
China Normal University, Guangzhou 510006, China}
\author{Zhen-Yu Wang}
\email{zhenyu.wang@m.scnu.edu.cn}

\affiliation{Guangdong Provincial Key Laboratory of Quantum Engineering and Quantum
Materials, School of Physics and Telecommunication Engineering, South
China Normal University, Guangzhou 510006, China}
\affiliation{Frontier Research Institute for Physics, South China Normal University,
Guangzhou 510006, China}

\begin{abstract}
We develop a framework to construct a new class of shortcuts to adiabaticity
for realization of adiabatic processes in short times. In contrast 
to previous strategies for accelerating adiabatic protocols, our approach speeds up 
the adiabatic process by applying the adiabatic Hamiltonian
at discrete points of the parameter space, thereby avoiding
the use of any counterdiabatic control that can be experimentally unfeasible,
skipping undesired points of the adiabatic path, and retaining the intrinsic robustness of adiabatic control.
We exemplify our general theory 
by construction of fast and noise-resilient control protocols for bosonic modes, three-level systems,
and interacting qubits.  Our framework offers a new route to design robust
and fast control methods for general quantum systems.
\end{abstract}
\maketitle

\paragraph{Introduction. ---}

Due to its intrinsic robustness against control imperfections, quantum
adiabatic control has a board range of applications from quantum
information processing to quantum sensing, such as stimulated Raman
adiabatic passage (STIRAP) for state transfer in three-level 
systems~\cite{vitanovStimulatedRamanAdiabatic2017,
vasilevOptimumPulseShapes2009,gaubatzPopulationTransferMolecular1990,
bergmannCoherentPopulationTransfer1998} and adiabatic quantum 
computing~\cite{Albash2018}. However, these quantum adiabatic control methods are
extremely slow following the traditional adiabatic condition. This slowness
can be a big problem because decoherence and noise could have enough time 
to spoil the desired states. 

\begin{figure}[b]
\centering \includegraphics[width=1.0\columnwidth]{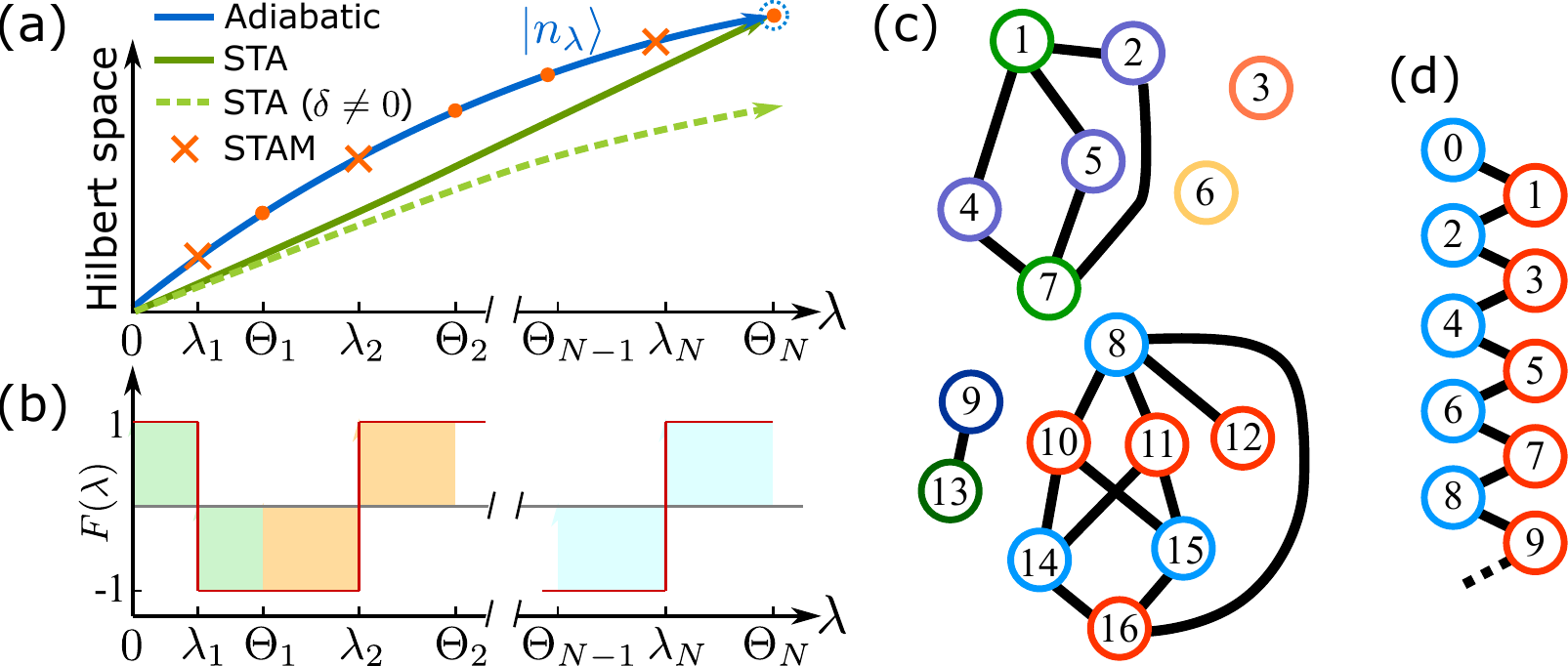} \caption{(a) Schematic trajectories (defined by the eigenstates of the total
Hamiltonian) to the target state (indicated by the blue dashed circle)
for an infinitely slow adiabatic process (solid blue), an ideal STA
(solid green), a STA with control errors (dashed light green), and
a STAM control (orange crosses). STA can be sensitive to control errors
$\delta$. STAM and adiabatic control are robust and have the same instantaneous
eigenstates of the total Hamiltonian. (b) A sketch of the function $F(\lambda)$.
The integral for the regions of the same color is zero. (c) Nodes
$n$ and $m$ that are not connected have $g_{n,m}=0$ in STAM. The
colors denote the types of the nodes. (d) An example of (c) for a
bosonic mode.}
\label{fig:Fig1}
\end{figure}

Shortcuts to adiabaticity (STA)~\cite{demirplak2003adiabatic,berry2009transitionless,chenShortcutAdiabaticPassage2010}
are alternative fast processes to realize the same final state evolution 
and have attracted much attention due to
practical and fundamental interests~\cite{torrontegui2013shortcuts,guery-odelinShortcutsAdiabaticityConcepts2019,del2019focus,santos2017generalized,hu2018experimental,santos2020optimizing,takahashi2019hamiltonian,abah2020quantum,kolbl2019initialization,chen2021shortcuts,zheng2022optimal}.
In existing STA methods, the total Hamiltonian $H^{\prime} = H+H_{{\rm CD}}$ is different from the original time-dependent Hamiltonian $H$
of adiabatic control. The addition of the counterdiabatic
Hamiltonian $H_{{\rm CD}}$ ensures that 
an eigenstate of $H(t=0)$ will evolve to the corresponding eigenstate of $H(T)$ at a later time $T$. 
However, implementation of $H_{{\rm CD}}$ can be challenging as it may require
forbidden transitions or unavailable experimental resources~\cite{torrontegui2013shortcuts}.
While for some cases it is possible to derive experimentally feasible STA control
using a suitable unitary transformation~\cite{PhysRevLett.109.100403}, the procedure
is complicated even for the case of two- and three-level 
systems~\cite{PhysRevLett.109.100403, bason2012high,PhysRevA.89.053408,PhysRevA.94.063411,baksicSpeedingAdiabaticQuantum2016,duExperimentalRealizationStimulated2016,zhouAcceleratedQuantumControl2017}.
Moreover, although a larger $H_{{\rm CD}}$ lessens the effects of dissipation
and decoherence by shortening the evolution time, $H_{{\rm CD}}$ generally adds additional
sources of control errors and changes the eigenstates of the total Hamiltonian $H^{\prime}$. The latter point implies
that the evolution of an initial ground state 
does not follow the ground-state evolution of the total Hamiltonian.  
Therefore, the robustness of STA is not guaranteed. 
Indeed, a recent comparative study~\cite{torosov2021coherent} on the control of two-level systems showed that
STA performs worse to typical control errors such as amplitude and detuning errors. See Fig.~\ref{fig:Fig1}(a).    
Designing STA protocols robust against control errors is a non-trivial
problem~\cite{song2021robust,ruschhaupt2012optimally,chenShortcutAdiabaticPassage2010,zheng2022optimal}.
In Refs.~\cite{xuBreakingQuantumAdiabatic2019,zheng2022accelerated}, it was found that
for the case of two-level systems one can, without the use of any $H_{{\rm CD}}$, speed up  
the quantum adiabatic evolution and retain the inherent robustness of adiabatic process,
by utilizing the necessary and sufficient quantum adiabatic condition~\cite{wangNecessarySufficientCondition2016}.
This shed new light on designing STA protocols in a new way for general
systems. 

In this Letter, we develop a general theory to construct a new type
of STA control. This STA by modulation (STAM) method is achieved by
dynamically modulating the functional forms of the parameters in the Hamiltonian for quantum
adiabatic evolution. The resulting protocols are simple, do not require additional
control, retain the robustness of adiabatic processes, and can avoid the control and interactions that are
not feasible in experiments. We illustrate the idea and show superior performances of STAM by applying it to bosonic systems, three-level systems, and interacting qubits.

\paragraph{STAM. ---}

We write the time-dependent Hamiltonian in its diagonal form
($\hbar=1$)
\begin{equation}
H(\lambda)=\sum_{n}E_{n}|n_{\lambda}\rangle\langle n_{\lambda}|,\label{eq:Ht}
\end{equation}
where $E_{n}$ are the eigenenergies, and the instantaneous eigenstates
$\{|n_{\lambda}\rangle\}$ depend on a monotonic function $\lambda=\lambda(t)$
which starts from an initial path point $\lambda(0)=0$. In the adiabatic
limit of an infinitely slow change
of $H$, the
time-ordered evolution $U=\mathcal{T}e^{-i\int_{0}^{t}Hdt^{\prime}}$
becomes the adiabatic evolution operator $U_{{\rm adia}}=\sum_{n}e^{-i\varphi_{n}}|n_{\lambda}\rangle\langle n_{0}|$.
Since the choice of gauge does not change the Hamiltonian and its evolution
operator~\cite{wangNecessarySufficientCondition2016}, for convenience we use the Born-Fock
gauge $g_{n,n}(\lambda)=0$ such that $\int_{0}^{\lambda}g_{n,n}(\lambda^{\prime})d\lambda^{\prime}=0$, where 
\begin{equation}
g_{n,m}(\lambda)\equiv\langle n_{\lambda}|i\frac{d}{d\lambda}|m_{\lambda}\rangle.\label{eq:gnm}
\end{equation}
In this gauge, $\varphi_{n}=\int_{0}^{t}E_{n}(t^{\prime})dt^{\prime}$ equals
the dynamic phase.  

To retain the intrinsic robustness of adiabatic process we keep using the Hamiltonian Eq.~(\ref{eq:Ht}). We realize the
target evolution $U_{{\rm adia}}$ within a finite time by choosing
a proper functional form of varying $H(\lambda)$. In the Supplemental Material \cite{SM} we obtain the fidelity \cite{Wang2008fid}
between the two evolution operators $U$ and $U_{{\rm adia}}$ as $\mathcal{F}=|{\rm Tr}U_{D}|/L_{{\rm dim}}\geq1-|D|$,
where the normalization factor $L_{{\rm dim}}={\rm Tr}I$ equals the
trace of the identity operator $I$; the unitary operator
\begin{equation}
U_{D}(\lambda)\equiv U_{{\rm adia}}^{\dagger}U=\mathcal{P}e^{i\int_{0}^{\lambda}W(\lambda^{\prime})d\lambda^{\prime}},\label{eq:UD_lambda}
\end{equation}
where $W(\lambda)=\sum_{n\neq m}e^{i[\varphi_{n}(\lambda)-\varphi_{m}(\lambda)]}g_{n,m}(\lambda)|n_{0}\rangle\langle m_{0}|$
and $\mathcal{P}$ indicates ordering with respect to $\lambda$;
the deviation $D$ from the target evolution has a universal upper bound
\begin{equation}
|D(\lambda)|<2\lambda L_{{\rm dim}}\sqrt{\epsilon_{{\rm ave}}\left(g_{{\rm max}}^{2}L_{{\rm dim}}+ g_{{\rm max}}^{\prime} 
 \right)g_{{\rm max}}},\label{eq:Dlambda}
\end{equation}
which is valid for any Hamiltonian.
Here the least upper bounds $g_{{\rm max}}=\sup_{{\rm \lambda^{\prime}\in(0,\lambda)}}|g_{n,m}(\lambda^{\prime})|$
and $g_{{\rm max}}^{\prime}=\sup_{{\rm \lambda^{\prime}\in(0,\lambda)}}|\frac{d}{d\lambda^{\prime}}g_{n,m}(\lambda^{\prime})|$
are fixed for a given evolution path. $|D(\lambda)|$ is small when
the factor $\epsilon_{{\rm ave}}=\sup_{{\rm \lambda^{\prime}\in(0,\lambda)}}|\int_{0}^{\lambda^{\prime}}e^{i[\varphi_{n}(\lambda^{\prime\prime})-\varphi_{m}(\lambda^{\prime\prime})]}d\lambda^{\prime\prime}|$
is small. Note that if $W(\lambda)$ commutes with itself at different
values of $\lambda$ and $\int_{0}^{\lambda}W(\lambda^{\prime})d\lambda^{\prime}=0$,
Eq.~(\ref{eq:UD_lambda}) gives $U_{D}(\lambda)=I$ and we realize
the target evolution $U_{{\rm adia}}$. 

To have $W(\lambda)$ commutes with itself, the first step is to choose 
\begin{equation}
|n_{\lambda}\rangle=e^{-iG\lambda}|n_{0}\rangle,\label{eq:State_n_lambda}
\end{equation}
for the eigenstates $\{|n_{\lambda}\rangle\}$ of the Hamiltonian
$H(\lambda)$ using a constant Hermitian operator $G$. In this manner
each $g_{n,m}=\langle n_{0}|e^{iG\lambda}i\frac{d}{d\lambda}e^{-iG\lambda}|m_{0}\rangle=\langle n_{0}|G|m_{0}\rangle$
becomes a constant, by using Eq.~(\ref{eq:gnm}). This also reduces the 
bound of the non-adiabatic error $|D|$ in Eq.~(\ref{eq:Dlambda}) since $g_{{\rm max}}^{\prime}=0$. 
The second step is to modulate the functional
forms of the dynamic phases with respect to $\lambda$ for all pairs
of $\{n,m\}$ with $g_{n,m}\neq0$ such that $e^{i[\varphi_{n}(\lambda)-\varphi_{m}(\lambda)]}=F(\lambda)$,
where $F(\lambda)\equiv e^{ij\pi}$ when $\lambda\in[\lambda_{j},\lambda_{j+1})$,
see Fig. \ref{fig:Fig1}(b). For other pairs of $\{n,m\}$ that $e^{i[\varphi_{n}(\lambda)-\varphi_{m}(\lambda)]}\neq F(\lambda)$,
we set $g_{n,m}=0$. In this way, $W(\lambda)=F(\lambda)G$ commutes
with itself, and from Eq.~(\ref{eq:UD_lambda}) $U_{D}(\Theta_{j})=e^{i\int_{0}^{\Theta_{j}}W(\lambda)d\lambda}=I$,
where $\Theta_{j}=2\sum_{k=1}^{j}(-1)^{j+k}\lambda_{k},\ (j=1,2,\ldots,N)$, see Fig.~\ref{fig:Fig1}(b).
Therefore, the controlled evolution
$U=U_{{\rm adia}}$ perfectly realizes the target evolution at $\Theta_{j}$. For simplicity we assume
equally spaced points $\lambda_{j}=\frac{\Theta_{N}}{2N}(2j-1)$ and
$\Theta_{j}=j\Theta_{N}/N$. 

While other ways are possible to achieve the desired functional forms
of $\{\varphi_{n}(\lambda)\}$ for $F(\lambda)$, here we apply a
sequence of control $H(\lambda)$ with $\lambda=\lambda_{1},\lambda_{2},\ldots,\lambda_{N}$.
In contrast to traditional adiabatic control where $H(\lambda)$
varies slowly over a continuous range of $\lambda$, applying
$H(\lambda)$ only at $\{\lambda_{j}\}$ can avoid unwanted points
$\lambda\notin\{\lambda_{j}\}$ that could be challenging or not feasible 
in experiments. We note that some of $\{|n_{\lambda}\rangle\}$ could
be chosen as dark states because only relative energies are important.

Any given initial and target states can be connected via Eq.~(\ref{eq:State_n_lambda})
with a suitable choice of $G$. This can be shown by an explicit
example that $G=\sum_{n=2}^{L_{{\rm dim}}}g_{n,1}|n_{0}\rangle\langle1_{0}|+{\rm h.c.}$.
In this case, we may choose $\varphi_{1}=E_{1}=0$ and, for $n\neq1$,
$E_{n}=E$, $\varphi_{n}(\lambda)=j\pi$ when $\lambda\in[\lambda_{j},\lambda_{j+1})$.
The sequence, where each $H(\lambda_{j})$ is applied for a pulse duration of $\pi/E$,
transforms an arbitrary initial state $|1_{0}\rangle$ to a general
superposition state $|1_{\Theta_{N}}\rangle=\cos(g\Theta_{N})|1_{0}\rangle+i\sin(g\Theta_{N})\sum_{n=2}^{L_{{\rm dim}}}\tilde{g}_{n,1}|n_{0}\rangle$
with $\tilde{g}_{n,1}=g_{n,1}/g$ and $g=\sqrt{\sum_{n=2}^{L_{{\rm dim}}}|g_{n,1}|^{2}}$. 

To find which $g_{n,m}$ can be nonzero, we consider the clusters where
the nodes marked by state labels $n$ and $m$ are connected (disconnected)
if $g_{n,m}\neq0$ ($g_{n,m}=0$), as illustrated in Fig. \ref{fig:Fig1}(c).
Since for $g_{n,m}\neq0$ the difference of dynamic phases $\varphi_{n}-\varphi_{m}$
has to be changed by an amount of $(2k_{nm}+1)\pi$ (with $k_{nm}$
being an integer) at each point $\lambda_{j}$, each cluster $\mathcal{C}$
of a size larger than one is required to have exactly two types
of nodes, $\mathcal{A}_{\mathcal{C}}$ and $\mathcal{B}_{\mathcal{C}}$.
If $n\in\mathcal{A}_{\mathcal{C}}$ and $m\in\mathcal{B}_{\mathcal{C}}$,
$e^{i[\varphi_{n}(\lambda)-\varphi_{m}(\lambda)]}=F(\lambda)$, while
for the pair $\{n,m\}\in\mathcal{A}_{\mathcal{C}}$ or $\{n,m\}\in\mathcal{B}_{\mathcal{C}}$
the nodes $n$ and $m$ cannot be linked (i.e., $g_{n,m}=0$). 

\paragraph{Fast preparation of coherent states.---}

A simple example of choosing the values of $g_{n,m}$ is illustrated
for a bosonic mode in Fig. \ref{fig:Fig1}(d), where we choose $g_{n+1,n}=i\alpha\sqrt{n+1}$
in the Fock state basis $\{|n\rangle\}$ ($n\geq0$).
Therefore $G=i\alpha a^{\dagger}-i\alpha^{*}a$, where $a$ is the
annihilation operator. $e^{-iG\lambda}$ is a displacement operator
which transforms the Hamiltonian $\omega a^{\dagger}a$ of a harmonic oscillator to $H(\lambda)=e^{-iG\lambda}\left(\omega a^{\dagger}a\right)e^{iG\lambda}=\omega a^{\dagger}a-\lambda\omega(\alpha a^{\dagger}+\alpha^{*}a)+\omega|\lambda\alpha|^{2}$.
Applying $H(1/2)$ for a duration of $\pi/\omega$
transfers the original ground state $|0\rangle$ to the ground state
of $H(1)$, i.e., a coherent state $|\alpha\rangle=e^{-iG}|0\rangle$. 
This is much faster than adiabatically varying the control $H(\lambda)$.

\begin{figure}[b]
\centering \includegraphics[width=\columnwidth]{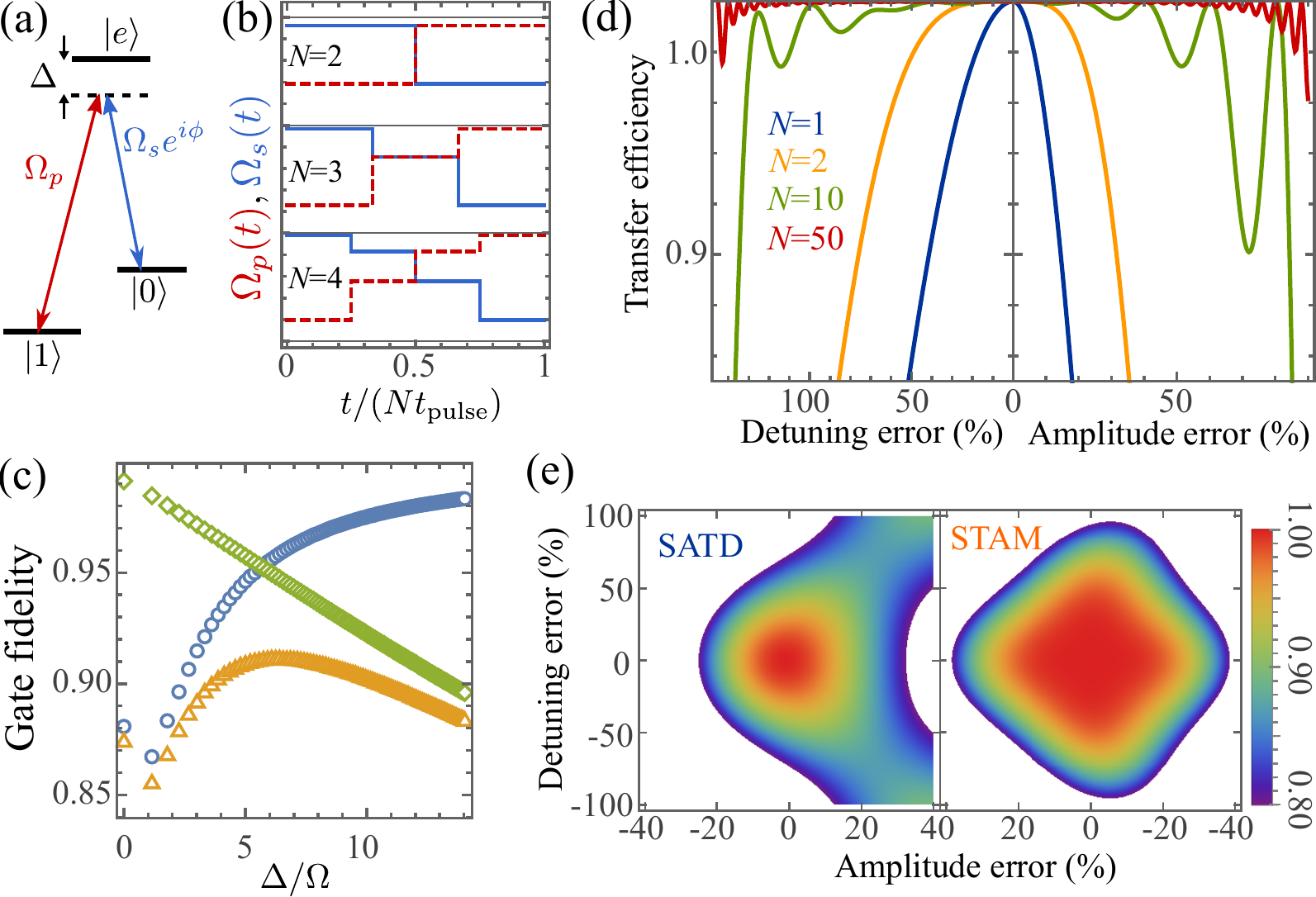} \caption{Adaptive control for three-level systems. (a) Indirect coupling of $|0\rangle$ and $|1\rangle$ qubit states
via two driving fields with a single-photon detuning $\Delta$ in
a $\Lambda$-type system. (b) The time envelopes of Stokes
$\Omega_{s}(t)$ (blue line) and pump $\Omega_{p}(t)$ (red dash)
pulses of the STAM protocol have simple pulse shapes. The cases of $N=2,3,4$ are shown.  (c) The gate fidelity
(yellow triangles) for $N=1$ and $\Theta_{N}=\frac{\pi}{2}$
in Eq.~(\ref{eq:U3gate}) within the qubit subspace at different detuning
$\Delta$, by including a relaxation rate $\Gamma_{e}=1.5\Omega/(2\pi)$
for decay of the intermediate state $|e\rangle$ into the qubit subspace
and a dephasing rate $\Gamma_{{\rm dep}}=0.05\Omega/(2\pi)$ on the
qubit states. Blue circles (green rectangles) are the result of setting
$\Gamma_{{\rm dep}}=0$ ($\Gamma_{e}=0$). (d) Robustness of the transfer
efficiency as a function of the amplitude and detuning errors (measured
in terms of $\Omega$) for STAM with different $N$.
(e) As (d) but for comparison between the
superadiabatic transitionless driving (SATD) developed in Refs. \cite{zhouAcceleratedQuantumControl2017,baksicSpeedingAdiabaticQuantum2016}
and the STAM. Both protocols have the same maximal value of control
field for $\Omega_{p}$ $\Omega_{s}$, and the same length of control
time $T=2\pi/\Omega$. See \cite{SM} for more details of the simulations.
}
\label{fig:Fig2}
\end{figure}

\paragraph{Intrinsic robustness.---}
A general, rigorous theory of the robustness for standard adiabatic control
can be seen from Eq.~(\ref{eq:Dlambda}). When the control becomes
adiabatic, $\epsilon_{{\rm ave}}$ is small (due to the averaging
by $e^{i[\varphi_{n}(\lambda)-\varphi_{m}(\lambda)]}$) and hence
the fidelity $\mathcal{F}$ is high \cite{wangNecessarySufficientCondition2016,xuBreakingQuantumAdiabatic2019}.
It is obvious that $\epsilon_{{\rm ave}}$ remains small even in the
presence of a positive error on $|\varphi_{n}(\lambda)-\varphi_{m}(\lambda)|$.
Thus a more adiabatic control is more robust against control
amplitude errors.   For STAM, a relative error $\delta_{e}$
between ${\varphi_{n}(\lambda)}$ induces $e^{i(\varphi_{n}-\varphi_{m})}=e^{ij\pi(1+\delta_{e})}$
when $\lambda\in[\lambda_{j},\lambda_{j+1})$. In~\cite{SM}, we
obtain $\epsilon_{{\rm ave}}=\frac{\Theta_{N}}{2N}[1+\pi|\delta_{e}|+O(\delta_{e}^{2})]$,
which remains small when $N$ is large. This implies that the robustness
is stronger when the distance $\lambda_{j+1}-\lambda_{j}$ between subsequent points is smaller.
Furthermore, the robustness of STAM is enhanced by the constraint
$g_{{\rm max}}^{\prime}=0$ because it gives a smaller deviation $|D(\lambda)|$.

For STAM there is another mechanism to enhance the robustness against
fast fluctuating errors. In contrast to traditional adiabatic approach that $\lambda$ changes in time, in STAM
 $H(\lambda)$ is applied at a fixed $\lambda=\lambda_{j}$ for a pulse duration $\tau_{p}$. At the same point $\lambda_{j}$, the random energy drifts $\beta(t)$ at different moments accumulate as a pulse area error $\delta=\int_{t_{j}}^{t_{j}+\tau_{p}}\beta(t^{\prime})dt^{\prime}$.
A fluctuating drift
$\beta(t)$ with a zero mean implies that the expectation value of $\delta$
is $\overline{\delta}=0$. This is the same mechanism of robustness
as the geometric gate in Ref. \cite{Zu2014Nature}, because fast
fluctuation at the boundary ($\beta$) hardly changes the area ($\delta$).
The variance $\overline{\delta^{2}}=\int_{t_{j}}^{t_{j}+\tau_{p}}\int_{t_{j}}^{t_{j}+\tau_{p}}dt_{1}dt_{2}\overline{\beta(t_{1})\beta(t_{2})}$
is negligible when the correlation time of the correlation function $\overline{\beta(t)\beta(0)}$
is small. 

\paragraph{Adaptive control for $\Lambda$-systems.--- }
To demonstrate STAM, consider a $\Lambda$-type three-level
system where there is no direct coupling between the two qubit states
$|0\rangle$ and $|1\rangle$, see Fig. \ref{fig:Fig2}(a). For this
system, the adiabatic jumping protocol in~\cite{xuBreakingQuantumAdiabatic2019}
and the STA protocols \cite{chenShortcutAdiabaticPassage2010} which
require direct coupling between $|0\rangle$ and $|1\rangle$ cannot
be used. To realize robust and fast state transfer between $|0\rangle$
and $|1\rangle$ by a dark state $|1_{\lambda}\rangle=e^{-iG\lambda}|1\rangle=\cos(\lambda)|1\rangle-\sin(\lambda)e^{i\phi}|0\rangle$
with $E_{1}=0$, we choose $g_{1,2}=i\cos\xi$, $g_{1,3}=i\sin\xi$,
$g_{2,3}=0$, and the initial eigenstates $|2_{\lambda=0}\rangle=\cos\xi e^{i\phi}|0\rangle-\sin\xi|e\rangle$,
$|3_{\lambda=0}\rangle=\sin\xi e^{i\phi}|0\rangle+\cos\xi|e\rangle$. See \cite{SM} for more details. 
The requirement $\langle0|H|1\rangle=0$ gives
$(E_{3}-E_{2})\cos(2\xi)=(E_{3}+E_{2})$ and the Hamiltonian of STAM,
$H(\lambda)=\Omega(|B_{\lambda}\rangle\langle e|+|e\rangle\langle B_{\lambda}|)+\Delta|e\rangle\langle e|,$
where the bright state $|B_{\lambda}\rangle=\sin\lambda|1\rangle+e^{i\phi}\cos\lambda|0\rangle$,
the detuning $\Delta=E_{2}+E_{3}$, and the overall coupling strength
$\Omega=\sqrt{-E_{2}E_{3}}$. To meet the condition for $F(\lambda)$,
$E_{2}=-(2k_{2}+1)\pi/t_{p}$ and $E_{3}=(2k_{3}+1)\pi/t_{p}$ with
the integers $k_{2},k_{3}\geq0$. 
The amplitudes of the Stokes ($\Omega_{s}=2\Omega\cos\lambda$) 
and pump ($\Omega_{p}=2\Omega\sin\lambda$)
pulses are illustrated in Fig. \ref{fig:Fig2}(b). 
Applying $H(\lambda)$ according
to the sequence $\lambda=\lambda_{1},\lambda_{2},\ldots,\lambda_{N}$
realizes a universal single-qubit quantum gate, 
\begin{equation}
U=\left(\begin{array}{cc}
\cos\Theta_{N} & (-1)^{N}e^{-i\phi}\sin\Theta_{N}\\
-e^{i\phi}\sin\Theta_{N} & (-1)^{N}\cos\Theta_{N}
\end{array}\right),\label{eq:U3gate}
\end{equation}
in the basis $\{|1\rangle,|0\rangle\}$ of the qubit subspace.
For even $N$ the gate is a rotation of $2\Theta_{N}$ along a direction in the
$x-y$ plane of the Bloch sphere, while for odd $N$ it is a $\pi$
rotation along an arbitrary direction. 


It is interesting that the value of $\Delta$ is adjustable for optimal
performance. For the resonant case $\Delta=0$, $t_{p}=\pi/\Omega$
reaches the quantum speed limit \cite{Pires2016,Marvian2016} and
the protocol for $N=1$ coincides with the geometric gate in Ref.~\cite{Zu2014Nature}.
On the other hand, a larger $\Delta$ can be used to reduce the effect of dissipation on the intermediate state $|e\rangle$.
The traditional method to suppress this dissipation is the adiabatic
elimination of $|e\rangle$ via a slow, second-order
control in the regime $\Omega\ll\Delta$. This slowness is a 
large obstacle in the presence of dephasing
noise since STIRAP is very sensitive to the two-photon detuning (equivalent
to dephasing noise on the qubit subspace)~\cite{guery-odelinShortcutsAdiabaticityConcepts2019}. 
In contrast, STAM can accelerate the control in the 
intermediate regime $\Delta\sim\Omega$ to overcome this obstacle, offering a higher gate fidelity in the presence of both dephasing
and dissipation {[}see Fig. \ref{fig:Fig2}(c){]}.  

Comparing with
existing robust STA for $\Lambda$ systems \cite{zhouAcceleratedQuantumControl2017,baksicSpeedingAdiabaticQuantum2016},
the robustness of STAM is much better and can be further enhanced by
using a larger $N$ {[}see Fig. \ref{fig:Fig2}(d),(e){]}.

\begin{figure}[b]
\centering \includegraphics[width=1\columnwidth]{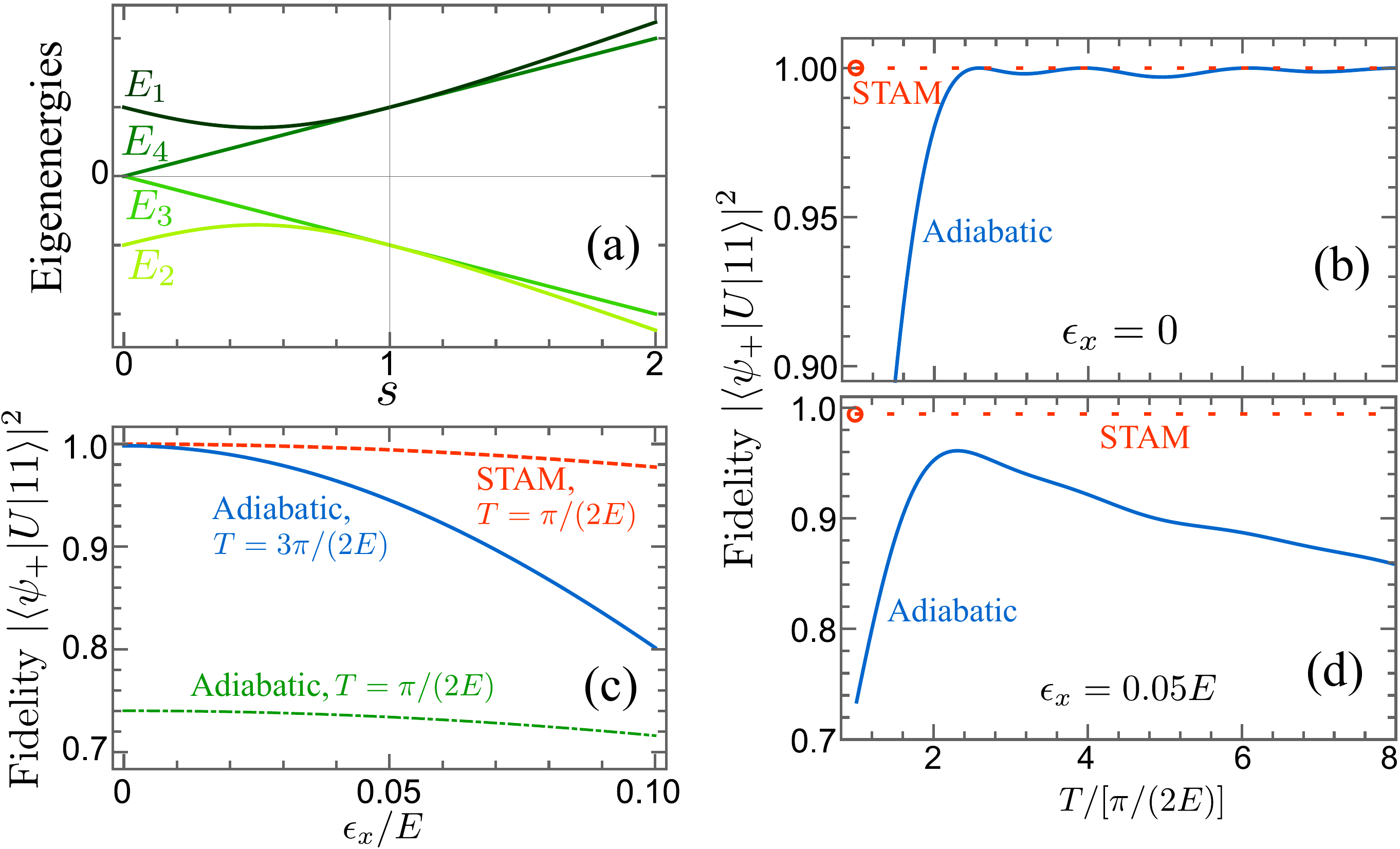} \caption{(a) The energy levels of Eq.~(\ref{eq:HSpinComputing}) are degenerate 
at the point $s=1$, where the target ground state encodes the solution. (b) The fidelity (blue solid line) of standard adiabatic
evolution to the target ground state at $s=1$ as a function of
the total evolution time $T$. The fidelity of the STAM protocol with $N=1$ is
shown by the orange dashed line and the minimal required time $T$
is indicated by an orange circle. (c) The fidelity as a function of
the strength of a local perturbation $\epsilon_{x}\sigma_{x}^{(1)}$
for standard adiabatic control (blue solid and green dash-dotted lines)
and STAM (orange dashed line). (d) As (b) but with the addition of
a weak local perturbation $0.05E\sigma_{x}^{(1)}$.}
\label{fig:Fig3}
\end{figure}

\paragraph{Application to coupled qubits.---}

As another example, consider the Hamiltonian for adiabatic quantum
computation, 
\begin{equation}
H=(1-s)H_{0}+sH_{1},\label{eq:HSpinComputing}
\end{equation}
where $H_{0}=-\frac{E}{2}(\sigma_{z}^{(1)}+\sigma_{z}^{(2)})$ and
$H_{1}=-E\sigma_{x}^{(1)}\sigma_{x}^{(2)}$ ($E>0$)~\cite{Xu2017}.
Starting from the ground state of $H_{0}$, $|11\rangle\equiv|1\rangle\otimes|1\rangle$,
slowly increasing the value of $s=s(t)$ from $s=0$ to $s=1$ is
supposed to adiabatically prepare the target solution state
$|\psi_{+}\rangle=\frac{1}{\sqrt{2}}(|00\rangle-|11\rangle)$. However,
$H$ has a gap closing at the final point $s=1$, with another
ground state $|\psi_{-}\rangle=\frac{1}{\sqrt{2}}(|01\rangle-|10\rangle)$, 
see Fig. \ref{fig:Fig3}(a). Therefore, a local fluctuation (e.g., a
weak perturbation $\epsilon_{x}\sigma_{x}^{(1)}$ with $\epsilon_{x}>0$)
breaks the parity symmetry of $H$ and leads to a wrong result
(e.g., $|\psi_{x}\rangle=\frac{1}{\sqrt{2}}(|\psi_{+}\rangle+|\psi_{-}\rangle$)
in the adiabatic limit, as shown in Fig.~\ref{fig:Fig3}. 

STAM avoids the use of the Hamiltonian at the degenerate point $s=1$ 
and is much faster. For this case we use $g_{1,2}=g_{2,1}=-1$
(other $g_{n,m}=0$) and $|2_{0}\rangle=i|00\rangle$ such that $|1_{\lambda}\rangle$
connects the initial ground state $|1_{0}\rangle=|11\rangle$ and
the target ground state $|\psi_{+}\rangle$. The other initial eigenstates
are the general superposition states $|3_{0}\rangle=\cos\beta e^{i\xi}|01\rangle-\sin\beta|10\rangle$
and $|4_{0}\rangle=\sin\beta e^{i\xi}|01\rangle+\cos\beta|10\rangle$.
Setting $\xi=0$, $\beta=\pi/4$, $E_{1}=-E_{2}=E$, $E_{3}=-E_{4}=E\sin(2\lambda)$,
we obtain the Hamiltonian $H(\lambda)=\cos(2\lambda)H_{0}+\sin(2\lambda)H_{1}$
as Eq.~(\ref{eq:HSpinComputing}) but with a different interpolation.
As shown in Fig.~\ref{fig:Fig3}, compared with standard adiabatic
control with a linear interpolation of $s=t/T$, STAM with $N=1$ 
achieves a higher state fidelity using a shorter time $\pi/(2E)$ 
and is much more robust to local fluctuations because the degenerate point $\lambda=\pi/4$ (i.e., $s=1$) is avoided. 

We can also use a different Hamiltonian, say, $H(\lambda)=\cot(2\lambda)H_{0}+H_{1}$,
which has similarities to the model of Landau-Zener-St\"{u}ckelberg transitions~\cite{shevchenko2010landau}. 
In contrast to the adiabatic control that continuously varies $\lambda$ from $\lambda=0$ to $\lambda=\pi/4$, our control
is bounded because the divergent
point $\lambda=0$ is avoided.  Because the interaction $H_{1}$ is a constant and 
$\cot(2\lambda)H_{0}$ can be  varied by 
the detuning of control field,
this control is faster and is easy to realize in systems such as a
nitrogen-vacancy (NV) center in diamond and nuclear spins
~\cite{taminiau2012detection, wang2016positioning, wang2017delayed, bradley2019ten}.   

\paragraph{Conclusion.---}

We developed a general theory for general quantum systems to construct
a new kind of STA protocols that speed up the adiabatic evolution only by the use of interactions 
in the original adiabatic Hamiltonian. Without the requirement of counterdiabatic
control or its equivalence using unitary transformations, the resulting STAM protocols
are more experimentally 
feasible and can avoid obstacles due to degenerate points, unbounded
values of control fields, and unavailable control resources. In addition to the ability to
retain the intrinsic robustness of quantum adiabatic control,
STAM protocols have other mechanisms to suppress the effects of
control errors and noise from the environment. 

Our results provide a new route to design robust and high-speed control methods, as demonstrated by the examples of preparation of coherent states, 
reliable evolution to the solution state in an adiabatic quantum computing model, as well as
a three-level system control protocol which has adjustable values of detuning for
optimal performance in the presence of both dephasing noise and excited
state decay. In future work, the pulse-sequence like STAM could be modified to have the 
ability of coherence protection as dynamical-decoupling (DD) pulse sequences~\cite{viola1999dynamical,yang2011preserving,xuBreakingQuantumAdiabatic2019}
and may be incorporated in DD-based sensing with adiabatic control~\cite{haase2018soft,whaites2022adiabatic}.

\begin{acknowledgments}
This work was supported by National Natural Science Foundation of
China (Grant No. 12074131) and the Natural Science Foundation of Guangdong Province (Grant No. 2021A1515012030). We thank Dr. Min Ming for useful suggestions and discussions. 
\end{acknowledgments}

%

\end{document}